\documentclass[twoside,11pt]{article}
\usepackage{jmlr2e_arXiv}
\usepackage{url}

\begin{document}

\title{\texttt{mlpy}: Machine Learning Python}

\author{\name Davide Albanese \email albanese@fbk.eu \\
        \addr Fondazione Bruno Kessler, Trento, Italy\\
	\name Roberto Visintainer \email visintainer@fbk.eu \\
        \addr Fondazione Bruno Kessler \& University of Trento, Italy\\	
	\name Stefano Merler \email merler@fbk.eu \\
        \name Samantha Riccadonna \email riccadonna@fbk.eu\\
	\name Giuseppe Jurman \email jurman@fbk.eu \\
        \name Cesare Furlanello \email furlan@fbk.eu \\
        \addr Fondazione Bruno Kessler, Trento, Italy}

\maketitle

\begin{abstract}
\texttt{mlpy} is a Python Open Source Machine Learning library built on top of NumPy/SciPy and the GNU Scientific Libraries.
\texttt{mlpy} provides a wide range of state-of-the-art machine learning methods for supervised and unsupervised problems and it is aimed at finding a reasonable compromise among modularity, maintainability, reproducibility, usability and efficiency.
\texttt{mlpy} is multiplatform, it works with Python 2 and 3 and it is distributed under GPL3 at the website \url{http://mlpy.fbk.eu}.
\end{abstract}
\section{Overview}
\label{sec:overview}
We introduce here \texttt{mlpy}, a library providing access to a wide spectrum of machine learning methods implemented in Python, which has proven to be an effective environment for building scientific oriented tools \citep{perez11python}. 
Although planned for general purpose applications, \texttt{mlpy} has the computational biology in general, and the functional genomics modeling in particular, as the elective application fields.
As a major applications example, we use \texttt{mlpy} methods to implement molecular profiling experiments that need to  warrant study reproducibility \citep{ioannidis09repeatability} and flawless results \citep{ambroise02selection}.
This task requires the availability of highly modular tools allowing the practioners to build an adequate workflow for the task at hand following authoritative guidelines \citep{maqc09maqcII}.
Such workflow involves a complex sequence of steps, both in the development and in the validation phases, starting from the upstream preprocessing algorithms to the downstream predictive analysis, repeated several times to accommodate the resampling schema.
The dimension of high-throughtput data involved (thousands of samples described by millions of features) and the large number of replicates needed to control bias effects make also efficiency an essential requirement.
\texttt{mlpy} is aimed at reaching a good compromise among code modularity, usability and efficiency.
In this spirit, \texttt{mlpy} finds a different equilibrium among all these characteristics, being more inclined towards flexibility than similar projects such as scikits-learn~\citep{pedregosa11scikit}, PyMVPA~\citep{hanke09pymvpa}, PyBrain~\citep{schaul10pybrain}, MDP~\citep{zito09mdp} and Shogun~\citep{sonnenburg10shogun}. 
In some areas the set of provided tools are among the most complete (\textit{e.g.}, wavelets) or even the only one (Canberra indicator for feature list stability) to be found.
In particular, \texttt{mlpy} supplies the biologist researcher with state-of-the-art implementations of many well known algorithms with attention for novel methods appearing in literature, and the bioinformatician more inclined to programming with a modular environment where to embed his favourite methods.
However, \texttt{mlpy} usage is not confined to bioinformatics: applications to computer vision, emotion detection, seismology, etology have been published in literature\footnote{\url{http://mlpy.sf.net/refs}}.
\texttt{mlpy} works on Python 2 and 3 and it is available for Linux, Mac OS X and Microsoft Windows (Xp, Vista, 7) platforms, under the GPL3 licence.
User documentation is written in Sphinx and it comes either online or as a downloadable manual in PDF format.
Together with the library description, a tutorial with several examples is provided as part of the documentation.
Due to design, separate documentation on API references is not needed: however, support for both final users and developers is offered by mean of a dedicated mailing list at the website \url{http://groups.google.com/group/mlpy-general}.
\texttt{mlpy} has been listed in the Machine Learning Open Source Software (MLOSS) repository\footnote{\url{http://mloss.org}} since February 2008.
\section{Background and Requirements}
\label{sec:tech}
\texttt{mlpy} is built on top of the NumPy/SciPy packages, the GNU Scientific Library (GSL) and it makes an extensive use of the Cython\footnote{\url{http://www.cython.org/}} language: these are prerequisites for the library installation.
NumPy and SciPy modules provide sophisticated $N$-dimensional array object, basic linear algebra functions and collect a variety of high level algorithms for science and engineering. 
The GNU Scientific Library (GSL) is the well-known module for numerical calculations written in C. 
Cython is a language very close to Python that allows generating very efficient C code and wrapping external C/C++ libraries.
\texttt{mlpy} includes an efficient Cython wrapper for the LibSVM \citep{chang11libsvm} and LibLinear \citep{fan08liblinear} C++ libraries. 
These implementations are reference for Support Vector Machines and large-scale linear classification, respectively.
\texttt{mlpy} is fully compatible with PyInstaller\footnote{\url{http://www.pyinstaller.org/}}, a software that converts Python packages and scripts into stand-alone executables for several platforms.
\section{Library Features}
\label{sec:feat}
The core of the library consists of a number of classical and more recent algorithms for classification, regression and dimensionality reduction, such as methods from the Support Vector Machines (SVM) and the Discriminant Analysis families, and their (mostly kernel-based) variants.
In particular, the set of classifiers includes Linear and Kernel SVM, Linear Discriminant Analysis (LDA), Diagonal Linear Discriminant Analysis (DLDA), Basic Perceptron, Logistic Regression, Elastic Net, Golub Classifier, Kernel Fisher Discriminant Analysis (KFDA), Parzen-based, k-nearest neighbor (KNN), classification tree, Maximum Likelihood.  
The implemented regressors are Ordinary (Linear) Least Squares, Linear and Kernel Ridge, Partial Least Squares, LARS, Elastic Net, Linear and Kernel SVM.  
Finally, Fisher Discriminant Analysis (FDA), Kernel FDA, Spectral Regression Discriminant Analysis (SRDA), Principal Component Analysis (PCA), Kernel PCA are the implemented dimensionality reduction algorithms.
Default values are provided for each classifier's parameter.
Distinct methods are deployed for the training (\texttt{learn()}), the testing (\texttt{pred()}) for classification and regression, and the projection (\texttt{transform()}) for the dimensionality reduction algorithms.
Whenever possible, functions are provided to access model parameters (for example, hyperplane coefficients or tranformation matrix) and other algorithm-specific information.
Kernel-based functions are managed through a common kernel layer.
In particular, the user can choose whether supplying either the data or a precomputed kernel in input space: linear, polynomial, Gaussian, exponential and sigmoid kernels are available as default choices, and custom kernels can be defined as well.
Many classification and regression algorithms are endowed with an internal feature ranking procedure: in alternative, \texttt{mlpy} implements the I-Relief algorithm. 
Recursive Feature Elimination (RFE) for linear classifiers and the KFDA-RFE algorithm are available for feature selection.
Methods for feature list analysis (for example the Canberra stability indicator \citep{jurman08algebraic}), data resampling and error evaluation are provided, together with different clustering analysis methods (Hierarchical, Memory-saving Hierarchical, k-means).
Finally, dedicated submodules are included for longitudinal data analysis through wavelet transform (Continuous, Discrete and Undecimated) and dynamic programming algorithms (Dynamic Time Warping and variants).
\paragraph{Example}
As a working example illustrating the use of the library in a simple machine learning task, we report the lines of code needed to perform a PCA followed by a SVM classification.
In particular, we detail the operational steps needed to project the samples of a UCI\footnote{\url{http://archive.ics.uci.edu/ml/datasets.html}} dataset on the cartesian plane generated by the first two principal components, to train a kernel SVM on the projected data and to test the trained model on the same data.
The dataset chosen for this dimensionality reduction example is the Iris dataset, collecting 150 observations of 3 different classes of iris flowers, each described by 4 attributes.
\begin{verbatim}
>>> iris.shape # 2d numpy array, 150 observations and 4 attributes
(150, 4)
>>> import mlpy # import the mlpy module
>>> pca = mlpy.PCA() # build a new PCA instance
>>> pca.learn(iris) # perform the PCA on the Iris dataset
>>> iris_pc = pca.transform(iris, k=2) # project Iris on the first 2 PCs
>>> svm = mlpy.LibSvm(kernel_type='linear') # build a new LibSVM instance
>>> svm.learn(iris_pc, labels) # train the model
>>> labels_pred = svm.pred(iris_pc) # test the model
>>> mlpy.error(labels, labels_pred) # compute the prediction error
0.033

\end{verbatim}
%
%
\acks{The authors acknowledge support from the EUP FP7 Project HiperDART and from the PAT funded Project ENVIROCHANGE.}
\vskip 0.2in
\bibliography{albanese12mlpy}
\end{document}